\documentclass[aps,prl,preprint,superscriptaddress,floatfix,showkeys]{revtex4-1}

\usepackage{rotating,graphicx}
\usepackage{dcolumn}
\usepackage{bm}
\usepackage{mathtools}

\begin{document}

\title{Unbound states in \bm{$^{17}\textrm{C}$}
    and \bm{$p$}-\bm{$sd$} shell-model interactions}


\author{S.~Kim \thanks{correspondingauthor}}
\email[]{sunjikim@ibs.re.kr}
\affiliation{Center for Exotic Nuclear Studies,
  Institute for Basic Science,
  Daejeon 34126, Republic of Korea}
\affiliation{Department of Physics and Astronomy, Seoul National University,
1 Gwanak-ro, Gwanak-gu, Seoul 08826, Republic of Korea}
\author{J.W.~Hwang}
\affiliation{Center for Exotic Nuclear Studies,
  Institute for Basic Science,
  Daejeon 34126, Republic of Korea}
\affiliation{Department of Physics and Astronomy, Seoul National University,
1 Gwanak-ro, Gwanak-gu, Seoul 08826, Republic of Korea}
\author{Y.~Satou}
\affiliation{Department of Physics and Astronomy, Seoul National University,
1 Gwanak-ro, Gwanak-gu, Seoul 08826, Republic of Korea}
\affiliation{
 Rare Isotope Science Project, Institute for Basic Science,
1 Gukjegwahak-ro, Yuseong-gu, Daejeon 34000, Republic of Korea
}
\author{N.A.~Orr}
\affiliation{
  LPC Caen,
  Normandie Universit\'{e},
  ENSICAEN, UNICAEN, 
  CNRS/IN2P3, F-14050 Caen, France}
\author{T.~Nakamura}
\author{Y.~Kondo}
\affiliation{Department of Physics, Tokyo Institute of Technology, 
2-12-1 O-Okayama, Meguro, Tokyo 152-8551, Japan}
\author{J.~Gibelin}
\author{N.L.~Achouri}
\affiliation{
  LPC Caen,
  Normandie Universit\'{e},
  ENSICAEN, UNICAEN, 
  CNRS/IN2P3, F-14050 Caen, France}
\author{T.~Aumann}
\affiliation{Institut f\"ur Kernphysik, Technische Universit\"{a}t Darmstadt, 
D-64289 Darmstadt, Germany}
\affiliation{ExtreMe Matter Institute EMMI and Research Division, 
GSI Helmholtzzentrum f\"ur Schwerionenforschung GmbH, D-64291 Darmstadt,
Germany}
\author{H.~Baba}
\affiliation{RIKEN Nishina Center, 
Hirosawa 2-1, Wako, Saitama 351-0198, Japan}
\author{F.~Delaunay}
\affiliation{
  LPC Caen,
  Normandie Universit\'{e},
  ENSICAEN, UNICAEN, 
  CNRS/IN2P3, F-14050 Caen, France}
\author{P.~Doornenbal}
\author{N.~Fukuda} 
\author{N.~Inabe}
\author{T.~Isobe}
\author{D.~Kameda}
\affiliation{RIKEN Nishina Center, 
Hirosawa 2-1, Wako, Saitama 351-0198, Japan}
\author{D.~Kanno}
\author{N.~Kobayashi}
\affiliation{Department of Physics, Tokyo Institute of Technology, 
2-12-1 O-Okayama, Meguro, Tokyo 152-8551, Japan}
\author{T.~Kobayashi}
\affiliation{Department of Physics, Tohoku University, Aoba, Sendai, 
Miyagi 980-8578, Japan}
\author{T.~Kubo}
\affiliation{RIKEN Nishina Center, 
Hirosawa 2-1, Wako, Saitama 351-0198, Japan}
\author{S.~Leblond}
\affiliation{
  LPC Caen,
  Normandie Universit\'{e},
  ENSICAEN, UNICAEN, 
  CNRS/IN2P3, F-14050 Caen, France}
\author{J.~Lee}
\altaffiliation{Present address: Department of Physics,
  University of Hong Kong, Pokfulam Road, Hong Kong.}
\affiliation{RIKEN Nishina Center, 
Hirosawa 2-1, Wako, Saitama 351-0198, Japan}
\author{F.M.~Marqu\'es}
\affiliation{
  LPC Caen,
  Normandie Universit\'{e},
  ENSICAEN, UNICAEN, 
  CNRS/IN2P3, F-14050 Caen, France}
\author{R.~Minakata}
\affiliation{Department of Physics, Tokyo Institute of Technology, 
2-12-1 O-Okayama, Meguro, Tokyo 152-8551, Japan}
\author{T.~Motobayashi}
\affiliation{RIKEN Nishina Center, 
Hirosawa 2-1, Wako, Saitama 351-0198, Japan}
\author{D.~Murai}
\affiliation{Department of Physics, Rikkyo University, 
3 Nishi-Ikebukuro, Toshima, Tokyo 171-8501, Japan}
\author{T.~Murakami}
\affiliation{Department of Physics, Kyoto University, 
Kyoto 606-8502, Japan}
\author{K.~Muto}
\affiliation{Department of Physics, Tohoku University, Aoba, Sendai, 
Miyagi 980-8578, Japan}
\author{T.~Nakashima}
\affiliation{Department of Physics, Tokyo Institute of Technology, 
2-12-1 O-Okayama, Meguro, Tokyo 152-8551, Japan}
\author{N.~Nakatsuka}
\affiliation{Department of Physics, Kyoto University, 
Kyoto 606-8502, Japan}
\author{A.~Navin}
\affiliation{GANIL, CEA/DRF-CNRS/IN2P3, 
F-14076 Caen Cedex 5, France}
\author{S.~Nishi}
\author{S.~Ogoshi}
\affiliation{Department of Physics, Tokyo Institute of Technology, 
2-12-1 O-Okayama, Meguro, Tokyo 152-8551, Japan}
\author{H.~Otsu}
\author{H.~Sato}
\author{Y.~Shimizu}
\author{H.~Suzuki}
\affiliation{RIKEN Nishina Center, 
Hirosawa 2-1, Wako, Saitama 351-0198, Japan}
\author{K.~Takahashi}
\affiliation{Department of Physics, Tohoku University, Aoba, Sendai, 
Miyagi 980-8578, Japan}
\author{H.~Takeda}
\author{S.~Takeuchi}
\affiliation{RIKEN Nishina Center, 
Hirosawa 2-1, Wako, Saitama 351-0198, Japan}
\author{R.~Tanaka}
\affiliation{Department of Physics, Tokyo Institute of Technology, 
2-12-1 O-Okayama, Meguro, Tokyo 152-8551, Japan}
\author{Y.~Togano}
\affiliation{ExtreMe Matter Institute EMMI and Research Division, 
GSI Helmholtzzentrum f\"ur Schwerionenforschung GmbH, D-64291 Darmstadt,
Germany}
\author{A.G.~Tuff}
\affiliation{Department of Physics, University of York, 
Heslington, York YO10 5DD, United Kingdom}
\author{M.~Vandebrouck}
\affiliation{Institut de Physique Nucl\'{e}aire, Universit\'e
Paris-Sud, IN2P3/CNRS, 91406 Orsay, France}
\author{K.~Yoneda}
\affiliation{RIKEN Nishina Center, 
Hirosawa 2-1, Wako, Saitama 351-0198, Japan}


\begin{abstract}

Unbound states in $^{17}{\rm C}$ were investigated 
via one-neutron removal from a $^{18}{\rm C}$ beam
at an energy of 245 MeV/nucleon on a carbon target.
The energy spectrum of $^{17}{\rm C}$, above the single-neutron decay threshold, was 
reconstructed using invariant mass spectroscopy from the measured momenta of the $^{16}{\rm C}$ fragment and neutron,
and was found to exhibit resonances at
$E_{\textrm{r}}$=0.52(2), 0.77(2), 1.36(1), 1.91(1), 2.22(3) and 3.20(1) MeV.
The resonance at $E_{\textrm{r}}$=0.77(2) MeV
[$E_x$=1.51(3) MeV] 
was provisionally assigned as the second $5/2^+$ state.
The two resonances at $E_{\textrm{r}}$=1.91(1) and 3.20(1) MeV 
[$E_x$=2.65(2) and 3.94(2) MeV] 
were identified, through comparison of the energies, cross sections and momentum distributions with shell-model and eikonal
reaction calculations, as $p$-shell hole states with spin-parities $1/2_1^-$ and $3/2_1^-$, respectively. 
A detailed comparison was made with the results obtained using a range of shell-model interactions.  The YSOX shell-model 
Hamiltonian, the cross-shell part of which is based on the monopole-based universal interaction,
was found to provide a very good description of the present results and those for the neighbouring odd-$A$ carbon isotopes -- in particular for the 
negative parity cross-shell states.
 
\end{abstract}


\keywords{One-neutron removal reaction, Neutron-rich carbon isotopes, Cross-shell states, Shell-model Hamiltonian}

\maketitle

Studies of nuclei away from $\beta$ stability
have revealed that shell structure
evolves as a function of the neutron ($N$) and
proton ($Z$) asymmetry.
Importantly, shell evolution is not only characterized by the appearance of new magic numbers and
the disappearance of conventional ones, but also by a variety of other features~\cite{Sorlin08,Otsuka18}.
Here the focus is placed on the lowest-lying cross-shell states in the neutron-rich isotopes of carbon.
In particular, by providing improved spectroscopic information on $^{17}{\rm C}$, the location of these
negative parity states is investigated in the light of specific components of shell-model effective interactions.
Such levels have to date been the subject of a relatively limited range of studies (see, for example, 
Refs.~\cite{Raimann96,Bohlen07,Ueno13,Ysatou14,Hwang17}).  Such states are formed by the creation of a hole in the $p$-shell orbitals --
$0\nu p_{1/2}$ or $0\nu p_{3/2}$ -- and their energies depend on the interplay between:
(1) the valence proton and neutron $T$=0 interaction,
(2) the valence neutron-neutron interaction and
(3) the valence neutron-hole interaction.
As such, the energy of these states, in particular across a range of isotopes,
enables elements of the shell-model effective interaction, which are not generally probed, to be explored.

In more general terms the neutron-rich carbon isotopes are of significant interest as they
form part of an isotopic chain that is experimentally accessible from the proton ($^{9}{\rm C}$)
to the neutron ($^{22}{\rm C}$) driplines.  Moreover they exhibit many of the phenomena
exhibited by light nuclear systems, ranging, for example, from single~\cite{Bazin95,Marques96,Nakamura99,Sauvan00,Maddalena01,Sauvan04,
Kobayashi12} and two-neutron haloes~\cite{Kobayashi12,Togano16} to large deformations \cite{Ogawa02,Ysatou08} and
retarded electromagnetic transition rates of excited states~\cite{Imai04,Ong08,DSuzuki08,Wiedeking08,Petri11}.
In parallel, effects such as those associated with the two-body interaction
have been studied~\cite{Yuan12b,Sieja11} owing to the relatively large magnitude of the residual interaction.  
In a more fundamental vein, $ab$-$initio$ structure calculations have also been applied to describe the neutron-rich carbon 
isotopes ~\cite{Petri12,Jansen14,Smalley15,Tran18}. 

Focusing on the energies of the first $2^+$ levels of the even-even neutron-rich carbon isotopes,
it has become clear that the vacancy (of two protons) in the proton $p$ shell
below the $Z$=8 shell closure leaves traces on the shell structure and inter-nucleon interaction for the valence neutrons.
First, the $2_1^+$ energies for $^{16,18,20}{\rm C}$ are very similar and quite low,
which has been interpreted as the disappearance of the $N$=14 sub-shell gap in carbon~\cite{Stanoiu08},
in contrast to the oxygen isotopes where the energy of the $2_1^+$ state rises markedly
at $^{22}{\rm O}$~\cite{Stanoiu04}. 
This is believed to be caused by the near degeneracy of the $\nu s_{1/2}$ and $\nu d_{5/2}$ orbits~\cite{Stanoiu08} in the carbon isotopes,
as confirmed by a recent single-neutron transfer reaction study of $^{17}{\rm C}$~\cite{Pereira20}.
Second, the conventional shell model utilizing, for example, 
the well established WBT interaction~\cite{Warburton92} does not reproduce the $2_1^+$ energies for $^{16,18,20}{\rm C}$ -- the 
calculated energies are $\sim$0.6 MeV higher than experiment~\cite{Stanoiu08,Kim18}.
Agreement can be obtained by reducing the neutron-neutron two-body matrix elements (MEs)
in the $sd$ shell by 25\% (WBT$^*$~\cite{Stanoiu08}), an effect which may arise from the loosely bound nature of the valence neutrons and/or the effect of core polarization~\cite{Kuo66}.
In this context, improved spectroscopy of neighbouring odd-$A$ neutron-rich
carbon isotopes is likely to be useful. 

Measurements of the magnetic dipole ($M1$) transition strengths
from the $1/2^+$ (212 keV) and $5/2^+$ (333 keV) excited states in $^{17}{\rm C}$ to the $3/2^+$ ground state
have revealed an anomalously retarded value for $B(M1:1/2^+$$\rightarrow$$3/2^+)$,
as compared to $B(M1:5/2^+$$\rightarrow$$3/2^+)$ ~\cite{DSuzuki08,Smalley15}. 
This has been explained~\cite{Suzuki08} by an enhancement in the tensor force in the $p$-$sd$ cross-shell interaction
in the SFO Hamiltonian~\cite{Suzuki03} -- specifically by replacing the relevant MEs of SFO
with those of the $\pi + \rho$ meson exchange tensor interaction of Ref.~\cite{Otsuka05}. 
We note that the more general issue of shell evolution driven by the tensor force 
was initially explored through systematic analyses of the effects of the monopole MEs~\cite{Otsuka18}. 
In this context, the YSOX shell-model interaction~\cite{Yuan12} 
has been developed using parameters from the SFO ($p$ shell), SDPF-M~\cite{Utsuno99} ($sd$ shell)
and the monopole-based universal interaction $V_{\textrm{MU}}$~\cite{Otsuka10} ($p$-$sd$ cross shell) 
that contains the central, $\pi + \rho$ tensor and spin-orbit force components.
The YSOX interaction has consequently been able to explain a variety of properties of light neutron-rich nuclei,
including the location of the neutron drip line~\cite{Yuan12}. 

The present study aimed at investigating excited states above the neutron-decay threshold
in the odd-$A$ carbon isotope $^{17}{\rm C}$.  
The high-energy single-neutron removal or ``knockout'' 
from a secondary $^{18}{\rm C}$ beam was employed in order to
populate 
single-particle states above the one-neutron separation energy,
including
the $p$-shell hole states -- spin-parity $J^{\pi}$=$1/2^-$ and $3/2^-$ -- 
candidates for which have been reported in a study of the $\beta$-delayed neutron emission from
$^{17}{\rm B}$~\cite{Ueno13}.  Employing high-energy neutron knockout provides, through the reconstructed 
momentum distribution of the $^{17}{\rm C}^*$ beam-like residue, a means to determine directly the orbital angular momentum
of the removed neutron and thus the parity of the state populated.
Furthermore, this approach allows spectroscopic factors to be deduced
which provide for a much more stringent test of shell-model interactions than
from the energies alone.
The results obtained, combined with earlier work, including that on the neighbouring odd-$A$ isotopes $^{15,19}{\rm C}$,
have been used to test a range of shell-model Hamiltonians. 

The secondary beam of $^{18}\textrm{C}$ was produced using the BigRIPS fragment separator~\cite{Kubo03}
at the RIKEN--RIBF laboratory~\cite{Yano07}.
A primary $^{48}\textrm{Ca}$ beam at 345 MeV/nucleon, and intensity of $\sim$80 pnA was used to bombard 
a 30-mm-thick beryllium production target. 
The $^{18}{\rm C}$ ions were transported to the SAMURAI facility~\cite{Kobayashi13} which was employed to undertake the 
measurements. 
The energy of the $^{18}{\rm C}$ beam at the mid-point of the secondary carbon reaction target
(1.8~g/cm$^2$ thick)
was 245 MeV/nucleon, with a momentum spread of $|\Delta p|/p$$\leq$3\%.
The intensity of the $^{18}{\rm C}$ beam
was around 2.3$\times$10$^3$ particles per second. 
Particle-identification of the beam ions  
was determined from the magnetic rigidity $(B\rho)$ (derived from a position measurement
at a dispersive focal plane of BigRIPS) together with 
the time-of-flight (TOF) and the energy loss $(\Delta E)$ measured using an ion chamber. 
The trajectory of the secondary beam onto the reaction target was deduced using two position sensitive drift chambers.
The beam velocity charged reaction products were momentum analysed
using the large-gap (80 cm) high acceptance superconducting dipole magnet of SAMURAI (central rigidity 7~Tm). 
The gap of the dipole was kept under vacuum using a chamber equipped
with thin large-area exit windows~\cite{Shimizu13} which minimized 
the amount of material encountered by both the fragments and neutrons. 
The trajectories of the charged fragments were determined using two drift chambers -- one
placed at the entrance and another at the exit of the dipole.  A 16-element plastic hodoscope, placed after the
second drift chamber, provided measurements of the $\Delta E$ and TOF with respect to a thin plastic start detector.

The beam velocity neutrons, emitted at forward angles, were detected using the multi-element NEBULA plastic scintillator
array~\cite{Nakamura16,Kondo20} placed $\sim$11 m downstream of the reaction target.
The array, which consisted of 120 neutron individual modules (12$\times$12$\times$180 cm$^3$)
and 24 charged particle veto modules, each with a thickness of 1~cm, was configured in two walls, each composed of two layers of 30 modules.
The NEBULA intrinsic detection efficiency of $31.6\pm 1.6$\% 
(for a 6 MeVee threshold setting)
was derived from the measurement of quasi mono-energetic neutrons produced using the $^7\textrm{Li}(p,n)$$^7\textrm{Be}$(g.s.+0.43 MeV) 
reaction at a proton energy of 250 MeV. 

The de-excitation $\gamma$ rays emitted from bound states of the charged fragments
were detected using the DALI2 array  which consisted of 140 NaI(Tl) scintillator detectors
surrounding the target~\cite{Takeuchi14,PieterPriv}. 
The array had a detection efficiency of 16(1)\% at 1~MeV
with a resolution of 150 keV (FWHM) after add-back analysis and Doppler correction. 

The invariant mass method was used to reconstruct the energy above the neutron-decay threshold in $^{17}\textrm{C}^*$.  
Specifically, the momentum vectors of the decay products, 
$(E_f,\bm{p}_f)$ and $(E_n,\bm{p}_n)$ for the fragment and neutron,
respectively,
were used to calculate the relative energy ($E_{\rm rel}$),

\begin{equation}
  E_{\rm rel}=\sqrt{(E_f+E_n)^2-|\bm{p}_f+\bm{p}_n|^2}-(M_f+M_n).
\end{equation}

Here $M_f(M_n)$ is the mass of the fragment (neutron). 
The excitation energy, $E_x$, is related to $E_{\rm rel}$ as $E_x$=$E_{\rm rel}$+$S_n$(+$E_{\gamma}$), 
where $S_n$ is
the one-neutron separation energy [$S_n$=0.735(18)~MeV for $^{17}\textrm{C}$~\cite{Wang12}]
and $E_{\gamma}$ is the energy of $\gamma$ rays emitted from the bound excited states of $^{16}\textrm{C}$, if populated.

\begin{figure}[!p]
  \resizebox{0.62\columnwidth}{!}{%
    \centering
    \includegraphics{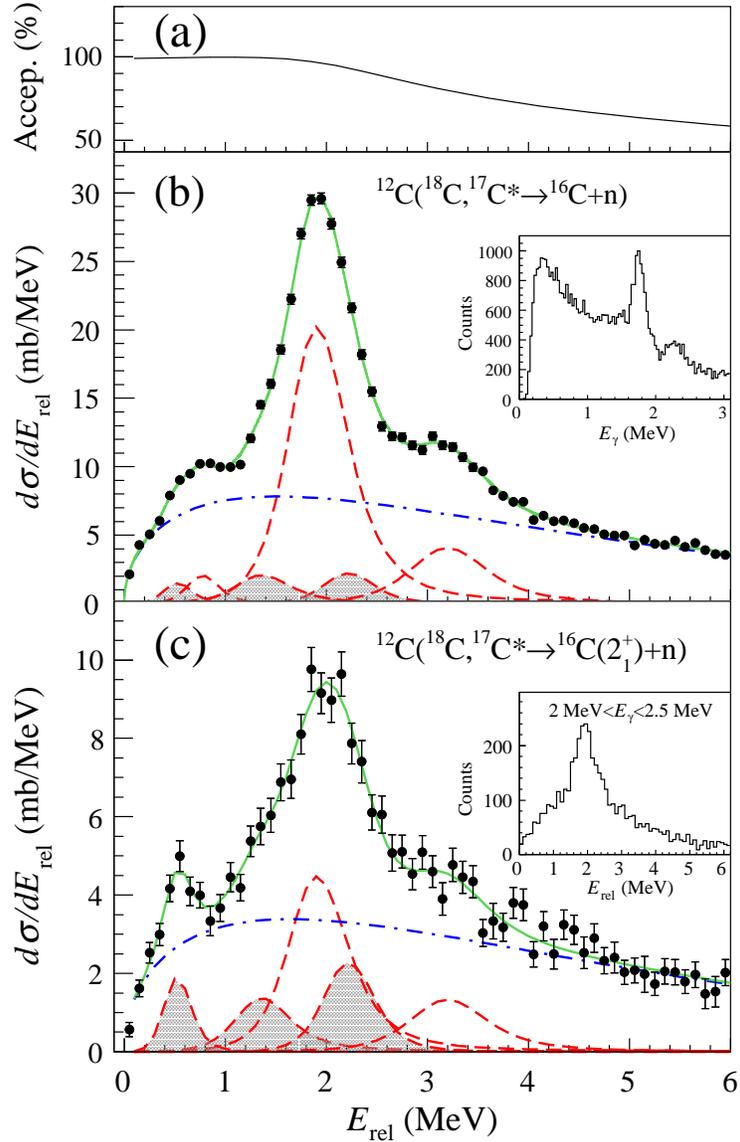}%
  }
  \caption{
    (a) The geometrical acceptance of NEBULA
    for one-neutron detection in coincidence with $^{16}\textrm{C}$
    for the $^{12}\textrm{C}(^{18}\textrm{C},$ $^{17}\textrm{C}^*)$ reaction
    at 245 MeV/nucleon. 
    (b) Relative energy spectrum
    of $^{17}\textrm{C}$ reconstructed from the measured momenta 
    of $^{16}\textrm{C}$ and a neutron. 
    (c) Relative energy spectrum obtained
    by requiring coincidence detection of the 1.77-MeV $\gamma$ ray
    from $^{16}\textrm{C}(2_1^+)$.
    The green solid lines
    represent the overall fit,
    the red dashed lines
    individual resonances 
    (the shaded peaks are those determined to be in coincidence with
    $^{16}\textrm{C}(2_1^+)$)
    and the blue dot-dashed lines the non-resonant continuum. 
    The inset in (b) displays the $\gamma$-ray energy spectrum 
    and that in (c) the relative energy spectrum (non-acceptance corrected)
    obtained by gating on the $\sim$2.3 MeV structure 
    in the $\gamma$-ray spectrum.
  }
\label{fig:spectrum_sub}
\end{figure}

In order to interpret the reconstructed $E_{\textrm{rel}}$ spectra a Monte Carlo simulation was developed which took into
account the geometry of the setup and detectors and their resolutions as well as the beam characteristics and target effects 
and the reaction itself (most notably through the momentum imparted to $^{17}\textrm{C}^*$ under the assumption of sudden removal of a neutron, which is subject to the Fermi motion, from the $^{18}\textrm{C}$ beam).  The geometrical acceptance as a function of $E_{\textrm{rel}}$ is shown in Fig.~\ref{fig:spectrum_sub}~(a).  The resolution in $E_{\textrm{rel}}$ was determined to scale
as $\Delta E_{\textrm{rel}}$$\approx$$0.4\sqrt{E_{\textrm{rel}}}$ MeV (FWHM).  As expected the primary effect on the resolution was the neutron detection (position and TOF).  In order to obtain the final fits (see below) to describe the
$E_{\textrm{rel}}$ spectra, the lineshapes of the different features (resonances and non-resonant continuum) were used as input for the simulations and the various parameters (resonance energies, widths, relative weights) were varied.  

The reconstructed $E_{\textrm{rel}}$ spectrum, obtained from the measured $^{16}{\rm C}$ and neutron coincidences,
is displayed in Fig.~\ref{fig:spectrum_sub}~(b) and is dominated by a clear resonance-like peak at 2~MeV straddled by two less 
prominent features at around 0.7 and 3.2~MeV along with a very broad underlying distribution.
The inset of panel (b) shows the coincident $\gamma$-ray spectrum which exhibits a strong peak from the well known
1.766(10)-MeV transition arising from the de-excitation of the $^{16}{\rm C}$ $2_1^+$ state~\cite{Tilley93}.  At slightly higher energy ($E_{\gamma}$$\sim$2.3~MeV) a much weaker and somewhat broader feature appears, which arises from the decay of some or all of the members of a triplet of levels located at around 4.1-MeV excitation energy~\cite{Maddalena01,Tilley93}.  We note that owing to the weakness of these transitions it was not possible to clearly associate them with any of the neutron decays observed here.
The spectra of panels (b) and (c) were obtained after subtracting away the contributions arising from reactions on materials other than the secondary target using data acquired with the target removed.

The spectrum displayed in Fig.~\ref{fig:spectrum_sub}~(c) shows the $E_{\textrm{rel}}$ spectrum when requiring a coincidence with the photopeak of the 1.77-MeV transition.  Whilst having a similar overall form to (b), this spectrum differs in its details.  Most notably, a clear relatively narrow peak appears at $E_{\textrm{rel}}$$\sim$0.5 MeV, which, in order to describe the broad feature at around 0.7~MeV in the inclusive spectrum, requires a $\gamma$-ray non-coincident strength
at $E_{\textrm{rel}}$$\sim$0.8~MeV [Fig.~\ref{fig:spectrum_sub} (b)] to also be present.  
The peak at $E_{\textrm{rel}}$$\sim$1.9 MeV in Fig.~\ref{fig:spectrum_sub} (c) is not only reduced in intensity but also
exhibits an asymmetric form with a width broader than in the inclusive spectrum Fig.~\ref{fig:spectrum_sub} (b) and requires the inclusion of $\gamma$-ray coincident strengths at $E_{\textrm{rel}}$$\sim$1.4 and 2.2 MeV.

The $E_{\textrm{rel}}$ spectra were fitted employing R-matrix~\cite{Lane58} line shapes  
with widths dependent on the decay energy and $\ell$-value of the neutron decay~\cite{Ysatou08} and a very broad underlying non-resonant continuum. 
The resonance energies ($E_{\textrm{r}}$), widths ($\mathit{\Gamma}$) and normalizations 
were determined using an iterative fitting procedure
applied to the inclusive [Fig.~\ref{fig:spectrum_sub} (b)] 
and the $\gamma$-ray coincidence [Fig.~\ref{fig:spectrum_sub} (c)] spectra, 
until good convergence was achieved in describing both. 
Here, the fit of the inclusive spectrum proceeded with six resonances,
while the $\gamma$-ray coincidence spectrum was fitted by five resonances
without the resonance at $E_{\textrm{rel}}$$\sim$0.8~MeV,
which presented too small an intensity to be included.
The non-resonant continuum was modelled using, as in earlier studies (e.g., Ref.~\cite{Ysatou08}), a distribution with the form,
$a\sqrt{E_{\textrm{rel}}}\exp(-bE_{\rm rel})$, with $a$ and $b$ taken to be fitting parameters. 

As the $\gamma$-ray spectrum exhibits a significant background, the analysis took into account
such components, including, for example,
the Compton scattering arising from the $\sim$2.3-MeV $\gamma$ rays,
that were expected to be included when gating on the 1.77-MeV photopeak.
Of the six resonances, those at $E_{\textrm{r}}$=0.52, 1.36 and 2.22~MeV could be clearly
determined to be in coincidence with the 1.77-MeV $\gamma$ ray (shaded peaks in Fig.~\ref{fig:spectrum_sub}). 
The peaks at $E_{\textrm{r}}$=1.91 and 3.20 MeV were determined not to be
in coincidence with the 1.77-MeV $\gamma$ ray. 
Even though they appear in Fig.~\ref{fig:spectrum_sub} (c), 
their counting rates were $\sim$3-4 times smaller than those
expected on the basis of the spectrum 
of Fig.~\ref{fig:spectrum_sub} (b)
given the NaI(Tl) array's detection efficiency
and the corresponding peaks thus arise from coincidences with
background $\gamma$ rays.

The resonance parameters ($E_{\textrm{r}}$ and $\mathit{\Gamma}$) for the 6 different neutron-decay transitions identified here, together with the associated partial cross sections ($\sigma_{-1n}^{\textrm{exp}}$), 
are summarized in Table~\ref{tab:resonance_parm}. 
The $E_x$ for the $E_{\textrm{r}}$=1.36 and 2.22-MeV resonances
are not listed since, as described below, they could not be unambiguously placed in the level scheme. 
The uncertainties quoted in the Table are those obtained by combining in quadrature 
the statistical and systematic contributions. 
The latter arises from the exact form of the non-resonant continuum,
the neutron detection efficiency and the geometrical acceptance correction. 
Specifically for $\sigma_{-1n}^{\textrm{exp}}$, the
errors resulting from statistical uncertainties are
14\%, 12\%, 12\%, 2\%, 14\% and 5\% 
for the $E_{\textrm{r}}$=0.52, 0.77, 1.36, 1.91, 2.22 and 3.20-MeV resonances,
respectively, 
while those from the systematic uncertainties 
for the same levels are 
25\%, 18\%, 42\%, 6\%, 35\% and 19\% 
arising from 
the choice of the non-resonant continuum, 
5\% (for all) from 
the neutron detection efficiency, and 2\% (for all) 
from the geometrical acceptance correction. 
The errors in $E_{\textrm{r}}$ and $\mathit{\Gamma}$ were dominated by the statistical uncertainties. 

Turning now to the spectroscopic factors, the single-neutron removal
cross section
$\sigma_{-1n}$
can be expressed in a factorized form as~\cite{Hansen03},
\begin{equation}
  \sigma_{-1n}=\sum_{n\ell j}\left(\frac{A}{A-1}\right)^N
  C^2S(J^{\pi},n\ell j)\sigma_{\textrm{sp}}(n\ell j,S_n^{\textrm{eff}}), 
\label{eqn:sigma_m1n}
\end{equation}
where $n\ell j$ 
refers to the quantum numbers of the removed neutron, 
$C^2S$ the spectroscopic factor, 
$\sigma_{\textrm{sp}}$ the single-particle cross section, 
$[A/(A-1)]^N$ the center-of-mass correction factor~\cite{Dieperink74}
with $A$ the mass number of the projectile
and $N$ the major oscillator quantum number ($N$=$2n$+$\ell$), 
and $S_n^{\textrm{eff}}$ the effective
single-neutron
separation energy
given by the sum of $S_n$ of the projectile
[$S_n(^{18}\textrm{C})$=4.18(4) MeV~\cite{Wang12}]
and $E_x$ of the state in $^{17}\textrm{C}^*$.

\begin{figure}[!p]
  \resizebox{0.95\columnwidth}{!}{%
    \centering
    \includegraphics{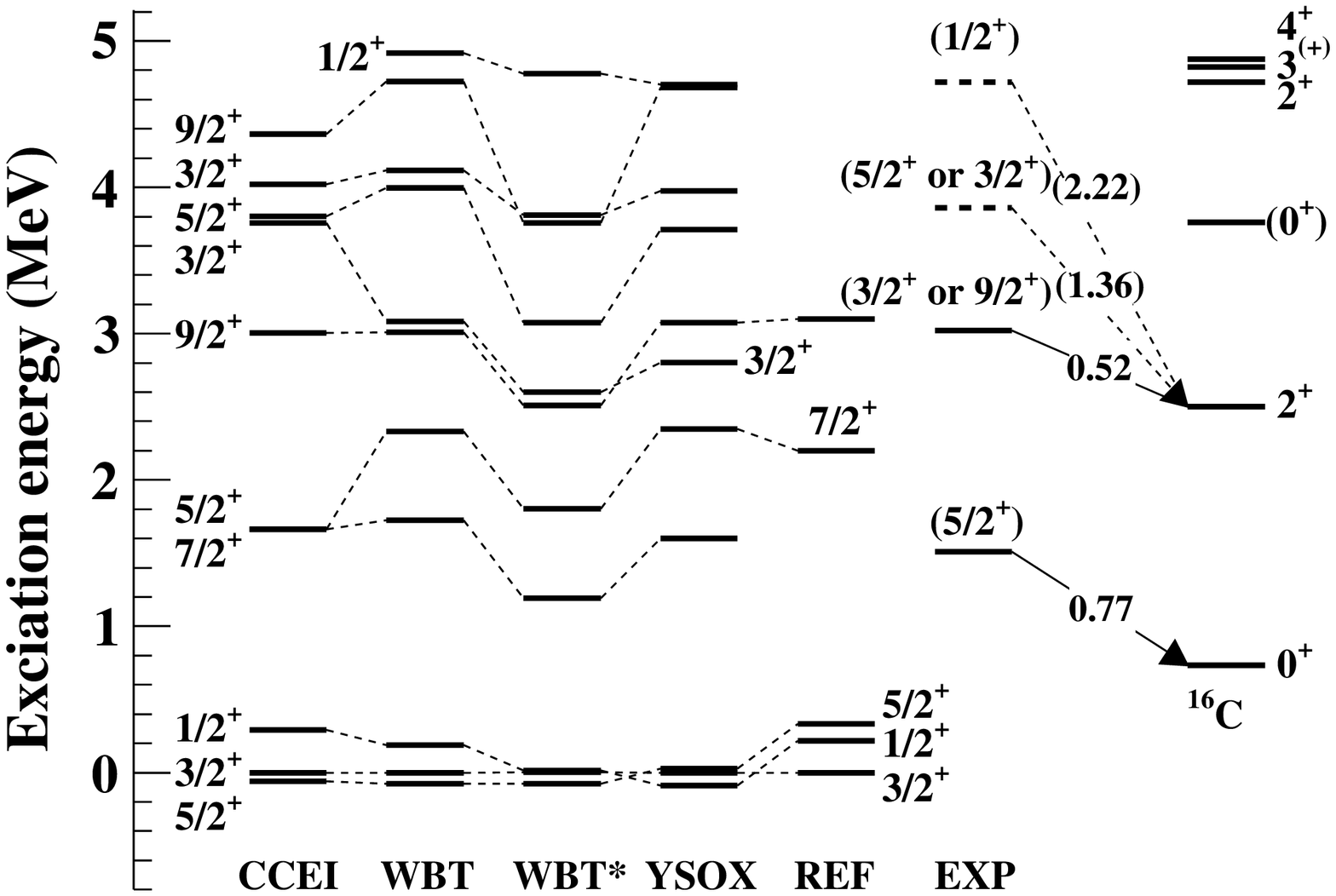}
  }
  \caption{Excitation energies of low-lying positive parity states
    of $^{17}{\rm C}$. 
    The energies calculated with the shell model
    using the CCEI~\cite{Jansen14}, WBT~\cite{Warburton92},
    WBT$^*$~\cite{Stanoiu08} and YSOX~\cite{Yuan12} interactions
    are compared with those
    of previously observed levels (REF)~\cite{Smalley15,Ysatou08,Bohlen07}
    and the present results (EXP).
    The excitation energies for theory are measured with respect to the $3/2_1^+$ state.
    Neutron decays leading to bound states in $^{16}{\rm C}$
    are shown by arrows
    with the energy of the transition indicated in MeV.
    Candidate levels for the 1.36
    and 2.22-MeV transitions are shown by dashed lines.
    \label{fig:spectroscopy_Sunji_17c_IV_pos}
  }
\end{figure}

\begin{figure}[!p]
  \resizebox{0.95\columnwidth}{!}{%
    \centering
    \includegraphics{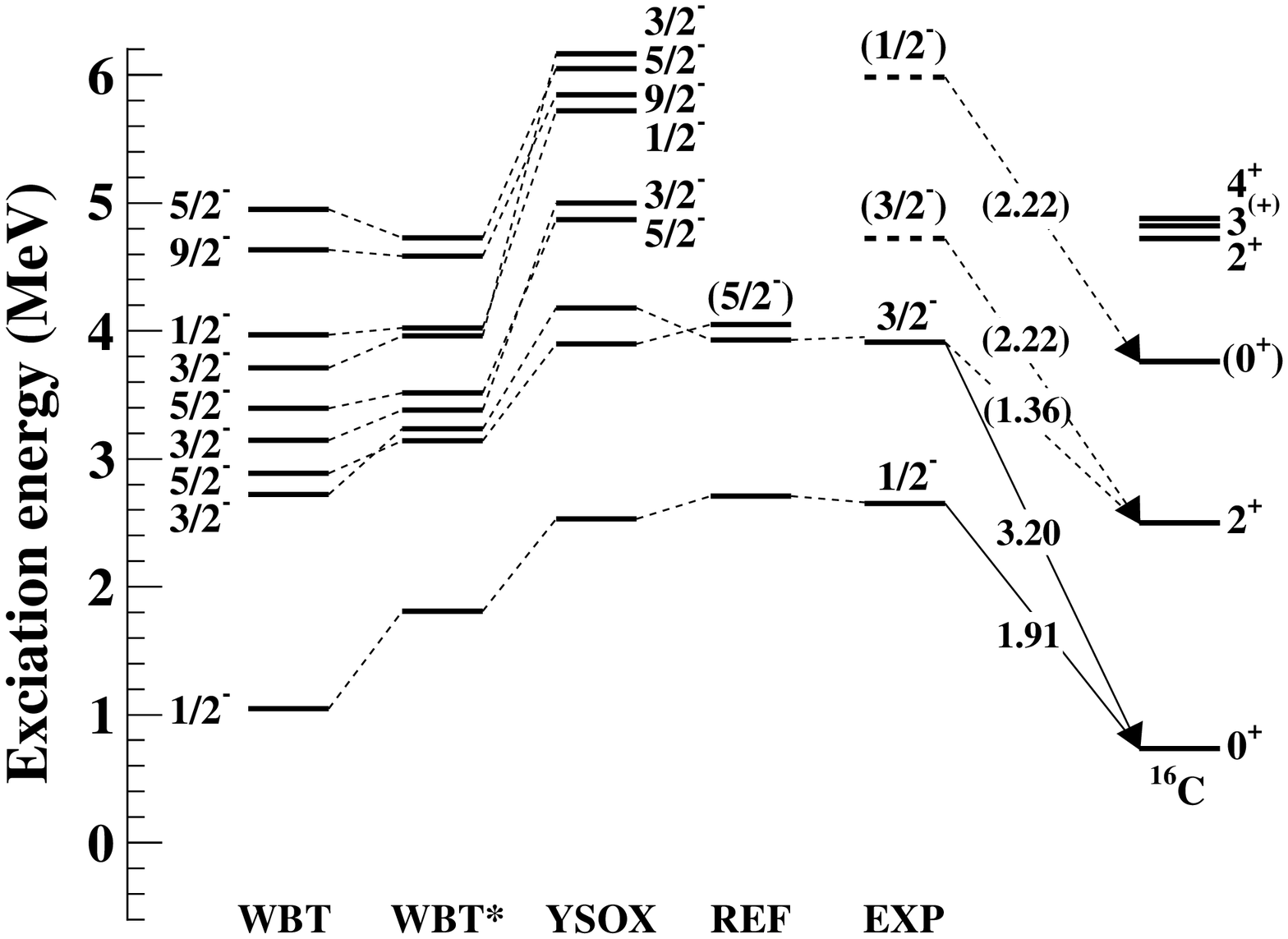}
  }
  \caption{Same as Fig.~\ref{fig:spectroscopy_Sunji_17c_IV_pos},
    but for negative parity states.
    \label{fig:spectroscopy_Sunji_17c_IV_neg}
  }
\end{figure}

Shell-model spectroscopic factors ($C^2S^{\textrm{th}}$)
and excitation energies ($E_{x}^{\textrm{th}}$)
were computed using the \textsc{nushellx@msu} code~\cite{NuShellX}
and the YSOX interaction~\cite{Yuan12}
in the $p$-$sd$ model space (Table~\ref{tab:resonance_parm}).
No explicit restriction in terms of the $\hbar \omega$ excitations
was applied. 
The shell-model spectroscopy for positive (negative) parity states
is compared with experimental data available for $^{17}{\rm C}$
in Fig.~\ref{fig:spectroscopy_Sunji_17c_IV_pos} 
(Fig.~\ref{fig:spectroscopy_Sunji_17c_IV_neg}), 
where the calculations are supplemented by the results
obtained using the WBT~\cite{Warburton92}, WBT$^*$~\cite{Stanoiu08}
and \textit{ab initio} 
Coupled-Cluster Effective Interaction 
(CCEI)~\cite{Jansen14} (for positive parity states only) interactions. 
The calculations utilizing the WBT and WBT$^*$ interactions 
were performed using
the \textsc{oxbash} code~\cite{OXBASH}
in the $s$-$p$-$sd$-$pf$ model space
for 0$\hbar \omega$ (1$\hbar \omega$) excitations
for the positive (negative) parity states. 

The $\sigma_{\textrm{sp}}$ 
was computed using \textsc{momdis}~\cite{Bertulani06}. 
The single-particle wave function 
was calculated using a Wood-Saxon potential
whose geometry is constrained to Hartree-Fock (HF) results
using the SkX interaction~\cite{Brown98} as described in Ref.~\cite{Gade08}. 
The range parameter of the nucleon-nucleon profile function~\cite{Ray79}
was fixed at 
zero for the present energy of 245 MeV/nucleon~\cite{Hansen03}.  
The nucleon density distribution of the $^{17}\textrm{C}$ core 
was estimated using a HF calculation
using the SkX interaction~\cite{Brown98}.
The density distribution of the carbon target
was chosen to be of a Gaussian form
with a point nucleon rms radius of 2.32~fm. 
An overall uncertainty of $\pm$15\% 
(included in the uncertainties assigned to the
$C^2S^{\textrm{exp}}$ 
in Table~\ref{tab:resonance_parm}) 
was estimated for $\sigma_{\textrm{sp}}$,
which arises from uncertainties 
in the size of the unbound core ($\pm$10\%) 
and in the reaction theory ($\pm$10\%)~\cite{Sauvan04,Carstoiu04}. 
The computed $\sigma_{\textrm{sp}}$ 
and associated $C^2S^{\textrm{exp}}$, 
deduced from $\sigma_{-1n}^{\textrm{exp}}$ using Eq.~(\ref{eqn:sigma_m1n}), 
for the relevant states are tabulated in Table~\ref{tab:resonance_parm}.  

\begin{figure}[!p]
  \resizebox{0.84\columnwidth}{!}{%
    \centering
    \includegraphics{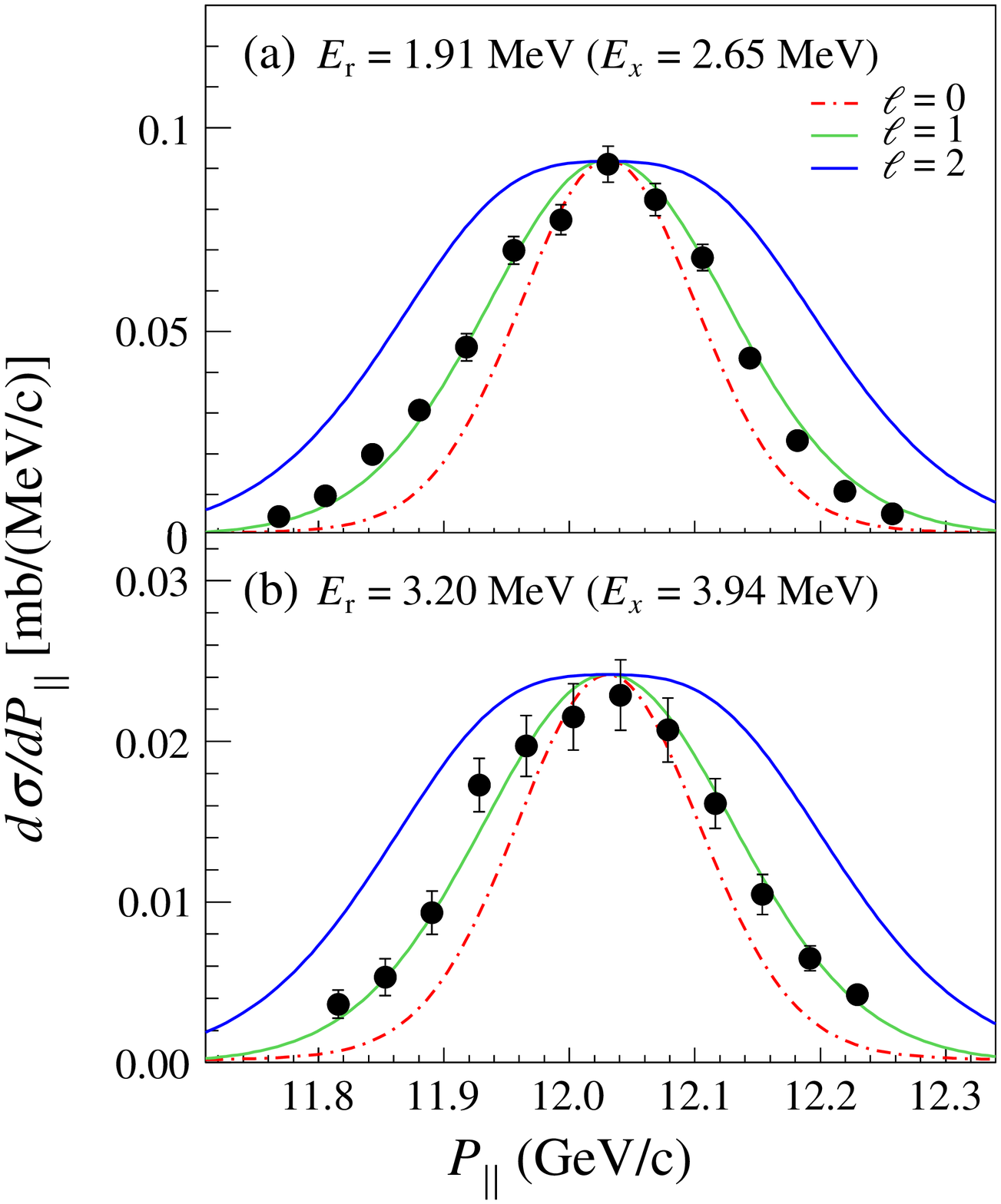}%
  }
  \caption{
    Experimental longitudinal momentum distributions (filled circles)
    for the $^{12}\textrm{C}(^{18}\textrm{C},$ $^{17}\textrm{C}^*)$ reaction 
    for (a) the $E_x$=2.65-MeV
    and (b) 3.94-MeV states.
    The theoretical distributions
    for removal of a neutron with $\ell$=0, 1 and 2
    (red dot-dashed, green solid and blue solid lines, respectively)
    include the experimental resolution. 
    \label{fig:ltmom1}
  }
\end{figure}

Longitudinal momentum distributions 
for the $^{17}\textrm{C}^*$ knockout residues 
populating the resonances observed at $E_{\textrm{r}}$=1.91 and 3.20~MeV 
[$E_x$=2.65(2) and 3.94(2) MeV]  
were deduced, and the results are shown in Fig.~\ref{fig:ltmom1}.
We note that for the other less strongly populated levels it was not possible 
to derive reliable results.
In order to construct the momentum distributions, an inclusive $E_{\textrm{rel}}$ spectrum was
created for each bin in longitudinal momentum and was fit as described above.  The error bars 
shown are statistical. 
Theoretical momentum distributions for neutron stripping,
which is the dominant mechanism at the present beam energy,
were calculated using the \textsc{momdis} code for removal of a valence neutron with
angular momentum of $\ell$=0, 1 and 2. 
In order to compare with the observed distributions, the predictions were convoluted with an experimental resolution of  
31 MeV/$c$ (sigma in the beam rest frame). 
As may be seen in Fig.~\ref{fig:ltmom1} 
both levels are clearly associated with the removal of an $\ell$=1
neutron.  As such, these are negative parity states formed by a neutron hole in the
$p$ shell.  
Considering the hierarchy of the neutron $p_{1/2}$ and $p_{3/2}$ orbits,
the former is closer to the Fermi level,
and the $J^{\pi}$ assignments of $1/2_1^-$ and $3/2_1^-$ for the $E_x$=2.65 and 
3.94-MeV levels, respectively, are in order.  These results are consistent with
the $\beta$-delayed neutron-decay experiment~\cite{Ueno13}
which observed the states in question at $E_x$=2.71(2) and 3.93(2) MeV 
and inferred the same assignments. 
As shown in Fig.~\ref{fig:spectroscopy_Sunji_17c_IV_neg} 
and in Table~\ref{tab:resonance_parm},
these conclusions are in very good agreement with the shell-model
calculations using the YSOX interaction. 
It is worthwhile noting that the use of neutron knockout,
as indicated in the introduction,
provided for more direct assignment.
For example, a $5/2^-$ assignment (possible in the $\beta$-decay study)
for the 3.93 MeV level,
is impossible in the present study
as there is no occupancy of the $p_{5/2}$ neutron orbital
in the $^{18}\textrm{C}$ projectile.

\begin{sidewaystable*}
  \caption{\label{tab:resonance_parm}
    Experimentally determined resonance energy $(E_{\textrm{r}})$, 
    excitation energy $(E_x)$
    and width $(\mathit{\Gamma})$ 
    of the unbound states in $^{17}\textrm{C}$
    produced via single-neutron removal from $^{18}\textrm{C}$
    in comparison with reaction and shell-model (YSOX~\cite{Yuan12})
    calculations.
    The theoretical excitation energy $(E_{x}^{\textrm{th}})$
    is given with respect to the $3/2_1^+$ state.}
\begin{ruledtabular}
  \begin{tabular}{dddcdddddc}
  \multicolumn{1}{c}{$E_{\textrm{r}}$ (MeV)} &
  \multicolumn{1}{c}{$E_{x}$ (MeV)} &
  \multicolumn{1}{c}{$\mathit{\Gamma}$ (MeV)} &
  \multicolumn{1}{c}{$\ell$ ($\hbar$)} &
  \multicolumn{1}{c}{$\sigma_{-1n}^{\textrm{exp}}$ (mb)} &
  \multicolumn{1}{c}{$\sigma_{\textrm{sp}}$ (mb)\footnotemark[1]$^,$\footnotemark[2]}&
  \multicolumn{1}{c}{$C^2S^{\textrm{exp}}$  \footnotemark[1]}&
  \multicolumn{1}{c}{$C^2S^{\textrm{th}}$} &
  \multicolumn{1}{c}{$E_{x}^{\textrm{th}}$ (MeV)} &
  $J^{\pi}$ \\ \hline
  0.77(2) &
  1.51(3) &
  \multicolumn{1}{c}{$\ll 0.36$\footnotemark[3]} &
  &
  0.87(24) & 
  21.93 &
  0.035(12) &
  0.015 &
  1.60 &
  ($5/2_2^+$) \\
  1.91(1) &
  2.65(2) &
  0.23(3) &
  1 &
  17.69(75) &
  20.26 &
  0.82(14) &
  1.350 &
  2.53 &
  $1/2_1^-$ \\
  0.52(2)\footnotemark[4] &
  3.02(4) &
  \multicolumn{1}{c}{$\ll 0.29$\footnotemark[3]} &
  &
  0.55(21) &
  &
  &
  &
  &
   \\
  3.20(1) &
  3.94(2) &
  0.32(9) &
  1 &
  4.61(50)\footnotemark[5] &
  19.40 &
  0.22(4)\footnotemark[5]&
  0.174 &
  4.18 &
  $3/2_1^-$ \\
  1.36(1)\footnotemark[4] &
  &
  0.27(13) &
  &
  1.73(44) \footnotemark[5] &
  &
  &
  &
  &
  \\
  2.22(3)\footnotemark[4] &
  &
  \multicolumn{1}{c}{$<$ 0.05} &
  &
  1.50(39) &
  &
  &
  &
  &
  \\
\end{tabular}
\end{ruledtabular}
\footnotetext[1]{An uncertainty of $\pm$15\% 
  associated with the reaction modeling
  is estimated for $\sigma_{\textrm{sp}}$
  and is included in the uncertainty in $C^2S^{\textrm{exp}}$ (see text).}
\footnotetext[2]{$S_n^{\textrm{eff}}$ derived from $E_x$
  were used in the reaction calculations.}
\footnotetext[3]{
  Upper limit corresponding to the experimental resolution at the corresponding relative energy.}
\footnotetext[4]{Observed in coincidence
    with $^{16}\textrm{C}(2_1^+)$ de-excitation $\gamma$ rays.}
\footnotetext[5]{If the $E_{\rm r}$=1.36-MeV transition 
  is a decay branch of the $3/2_1^-$ level to $^{16}{\rm C}(2_1^+)$, the total cross section 
  to the $3/2_1^-$ state is 6.34(67) mb and $C^2S^{\textrm{exp}}$=0.31(6) (see text).}
\end{sidewaystable*}

The resonance at $E_{\textrm{r}}$=0.77 MeV was not found to be 
associated with $\gamma$-ray emission, 
and is thus located at $E_{x}$=1.51(3) MeV. 
The YSOX interaction shell-model predictions place two positive parity states, 
$5/2_2^+$ and $7/2_1^+$ (Fig.~\ref{fig:spectroscopy_Sunji_17c_IV_pos}), 
above the single-neutron emission threshold of $^{17}{\rm C}$
and below the $^{16}{\rm C}(2_1^+)$+$n$ threshold of $E_x$=2.50 MeV. 
A $J^{\pi}$=$7/2_1^+$ assignment has been reported 
for a level at $E_x$=2.20 MeV 
observed in the $(p,p')$ reaction~\cite{Ysatou08}.  In addition, direct single-neutron removal from 
$^{18}{\rm C}$ is not expected to populate the $7/2_1^+$ state.
As such the $E_x$=1.51-MeV state we observe here is very likely the $5/2_2^+$ level, which is
predicted to lie at $E_{x}^{\textrm{th}}$=1.60 MeV by the YSOX calculations (Table~\ref{tab:resonance_parm}).  
We note that an importance-truncated no-core shell-model (IT-NCSM) calculation
incorporating the chiral nucleon-nucleon and three-nucleon interactions  
predicts the $5/2_2^+$ state at 1.78 MeV when employing the largest tractable basis size ($N_{\textrm{max}}$=6)~\cite{Smalley15}. 
In addition, the predicted energy difference between the $5/2_1^+$ and $5/2_2^+$ levels of 
1.27 MeV is compatible with the present observation of 1.18(3) MeV.

The $E_{\textrm{r}}$=0.52-MeV resonance
was observed in coincidence
with the $^{16}{\rm C}(2_1^+)$ de-excitation $\gamma$ ray
and exhibits the lowest decay energy 
of all the states observed here. 
Since the phase space for the two-body decay
is proportional to the square root of the decay energy,
such a small transition energy is very probably associated
with a state which has a unique decay path. 
Further the $E_{\textrm{rel}}$ spectrum obtained 
by gating on the $\sim$2.3 MeV structure
in the $\gamma$-ray spectrum,
which corresponds to a multiplet
of states at around $E_x$=4.0 MeV in $^{16}{\rm C}$,
did not exhibit
any enhanced strength around $E_{\textrm{rel}}$=0.52 MeV 
[Fig.~\ref{fig:spectrum_sub}~(c) inset].
As such, the state in question is
located at $E_x$=3.02(4) MeV. 
Two of the YSOX interaction shell-model levels -- $3/2_2^+$ and $9/2_1^+$  --
which are predicted to lie at $E_{x}^{\textrm{th}}$ of 2.80 and 3.07 MeV, respectively,
are possible candidates for this state (Fig.~\ref{fig:spectroscopy_Sunji_17c_IV_pos}). 
We note that a candidate for the $9/2_1^+$ level was 
identified at 3.10(2) MeV in the three-neutron transfer reaction
$^{14}{\rm C}$($^{12}{\rm C},$ $^9{\rm C})$~\cite{Bohlen07}. 
The decay of the $3/2_2^+$ level proceeds via $s$-wave neutron emission,
while that of $9/2_1^+$ via $d$-wave neutron emission.
The former is expected to result in a very broad structure, whilst the higher angular momentum of the latter 
would result in a rather narrow peak, as observed here, and is thus favoured.
Such an assignment would imply, however, that the single-neutron knockout from $^{18}{\rm C}$ populating
this level proceeds via a multi-step process. For example, in a first step the $^{18}{\rm C}(5_1^+)$ level
at $E_x^{\textrm{th}}$=7.68 MeV (YSOX) might be populated by inelastic scattering and subsequently decay via 
$s$-wave neutron emission to the $^{17}{\rm C}(9/2_1^+)$ state.

The remaining two peaks observed at $E_{\textrm{r}}$=1.36 and 2.22 MeV 
were populated with cross sections
of the order of 1 mb, and both were observed in coincidence with 
the de-excitation $\gamma$ ray from $^{16}\textrm{C}(2_1^+)$. 
In order to explore possible $J^{\pi}$ assignments and locate the states in the energy level scheme of $^{17}{\rm C}$ 
the properties of the single-neutron decay of low-lying shell-model (YSOX) levels have been explored through calculations of
the partial decay widths and branching ratios.  Although no definite conclusion could be reached, in particular because of the increased level densities at higher $E_x$, tentative suggestions for their
possible placements are indicated in Figs.~\ref{fig:spectroscopy_Sunji_17c_IV_pos} and
\ref{fig:spectroscopy_Sunji_17c_IV_neg}. 
It may be noted that if, as shown in Fig.~\ref{fig:spectroscopy_Sunji_17c_IV_neg}, the $E_{\textrm{r}}$=1.36-MeV resonance
corresponds to a decay branch of the $3/2_1^-$ state to $^{16}{\rm C}(2_1^+)$,
the total neutron removal cross section will be
$\sigma_{-1n}^{\textrm{exp}}(3/2_1^-)$=6.34(67) mb with a corresponding spectroscopic factor of 
$C^2S^{\textrm{exp}}$=0.31(6).

\begin{figure}[!p]
  \resizebox{0.95\columnwidth}{!}{%
    \centering
    \includegraphics{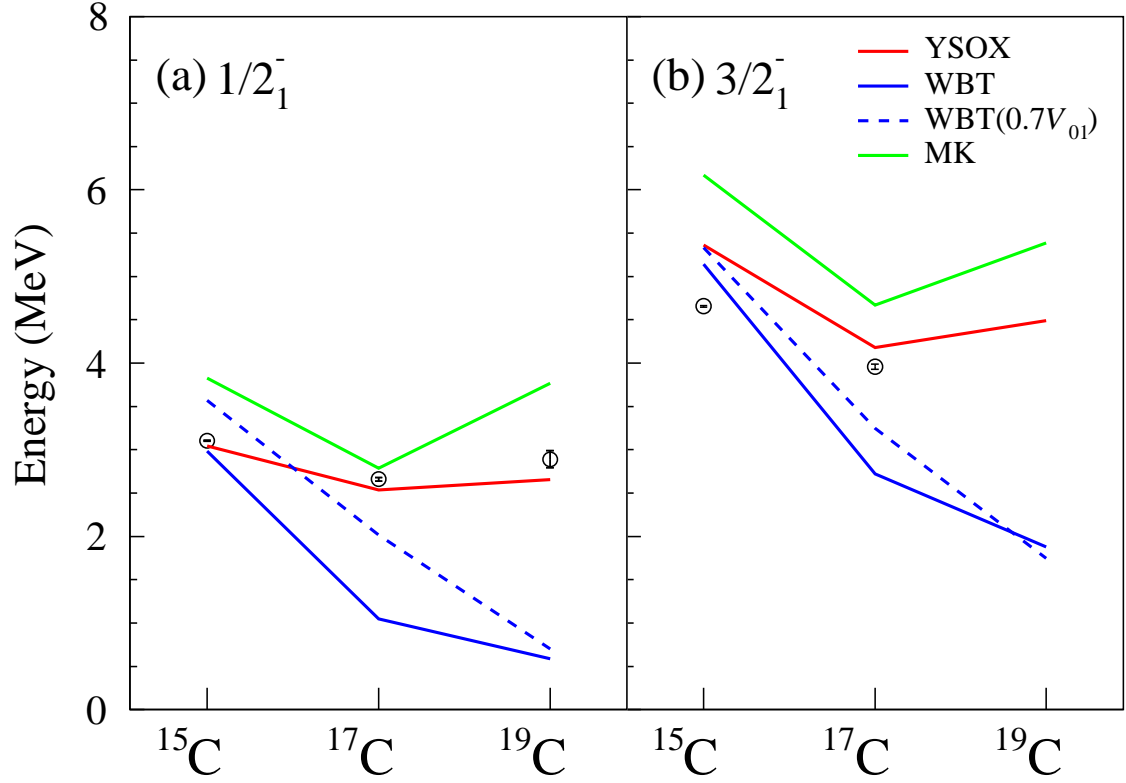}%
    }
  \caption{
    Energies with respect to 
    the $1/2_1^+$ ($^{15}{\rm C}$), $3/2_1^+$ ($^{17}{\rm C}$) 
    and $1/2_1^+$ ($^{19}{\rm C}$) states
    of (a) the $1/2_1^-$
    and (b) $3/2_1^-$ states in $^{15,17,19}{\rm C}$.
    The experimental energies (open circles) are from the present work for $^{17}{\rm C}$ 
    and Refs.~\cite{Ajzenberg-Selove91,Hwang17} for $^{15,19}{\rm C}$. 
    Shell-model calculations utilizing the
    YSOX~\cite{Yuan12} (red solid lines),
    WBT~\cite{Warburton92} (blue solid lines),
    WBT(0.7$V_{01}$)~\cite{Ueno13} (blue dashed lines) and
    MK~\cite{Millener75} (green solid lines) interactions
    are shown for comparison (see text).
  }
\label{fig:migration_1hm_3hm_IIIb}
\end{figure}

Turning to the negative parity cross-shell states, Ueno \textit{et al.}~\cite{Ueno13} 
have argued that the lower than observed energies 
predicted by the WBT shell-model interaction for the lowest-lying negative parity
$1/2_1^-$ and $3/2_1^-$ states in $^{17}{\rm C}$ 
(Fig.~\ref{fig:spectroscopy_Sunji_17c_IV_neg}) 
can be remedied by reducing by 30\% the diagonal pairing terms
of the $sd$-shell MEs, $V_{J=0,T=1}$ 
[WBT(0.7$V_{01}$)]. 
Here $J$ and $T$ refer to the angular momentum and isospin, 
respectively, of the corresponding two-particle state.
Reduced effects of polarization of the core (carbon isotopes having two less protons in the $p$ shell than oxygen)
are the source of the diminished pairing terms in this model.
This approach is examined in the following by comparing the present results and those for the 
neighbouring odd carbon isotopes, $^{15,19}{\rm C}$, with prediction using other shell-model Hamiltonians.

Fig.~\ref{fig:migration_1hm_3hm_IIIb}
compares the energies with respect to the $1/2_1^+$ ($^{15}{\rm C}$), $3/2_1^+$ ($^{17}{\rm C}$) 
and $1/2_1^+$ ($^{19}{\rm C}$) states
of the (a) $1/2_1^-$ 
and (b) $3/2_1^-$ states in $^{15,17,19}{\rm C}$ 
with shell-model calculations utilizing the YSOX~\cite{Yuan12} (described earlier), WBT~\cite{Warburton92}, 
WBT(0.7$V_{01}$)~\cite{Ueno13} and 
MK~\cite{Millener75} interactions. 
A 0,2$\hbar \omega$ (1,3$\hbar \omega$) basis was used 
for the \textsc{oxbash} calculations 
of the positive (negative) parity states in $^{15}{\rm C}$. 
Calculations with the MK interaction were performed in the $p$-$sd$ model space
using the \textsc{oxbash} code. 
As pointed out in Ref.~\cite{Ueno13}, 
the WBT energy for $1/2_1^-$ in $^{17}{\rm C}$ 
is too low by $\sim$1.5 MeV as compared to experiment
[$E_x$=2.65(2) MeV], 
while the energy calculated using the WBT(0.7$V_{01}$) interaction 
is in much better agreement. 
For the $1/2_1^-$ level in $^{19}{\rm C}$ [$E_x$=2.89(10) MeV~\cite{Hwang17}],
the WBT prediction is again too low ($E_x^{\textrm{th}}$=0.64 MeV),
while the WBT(0.7$V_{01}$) does not provide a better estimate either.
This difference in the effects of reducing the diagonal pairing terms
of the WBT interaction on the calculated energies of the $1/2_1^-$ states in $^{17,19}{\rm C}$
can be understood as follows. 
The dominant neutron configurations for the $3/2_1^+$ ground state of $^{17}{\rm C}$ 
are $\nu(d_{5/2})^3$ (seniority $v=3$) and $\nu(d_{5/2})^2(s_{1/2})^1$~\cite{Yuan12,Maddalena01} 
and for the $1/2_1^-$ level $\nu(p_{1/2})^{-1}(sd)^4$. 
The reduced pairing has little effect on the binding of the ground state,
while it makes the $1/2_1^-$ state less bound,
resulting in a higher $E_x$. 
In $^{19}{\rm C}$, 
the $1/2_1^+$ ground state~\cite{Bazin95,Nakamura99,Maddalena01}
and the $1/2_1^-$ state
have primarily $\nu(s_{1/2})^1(sd)^4$ and $\nu(p_{1/2})^{-1}(sd)^6$ neutron configurations, 
respectively.
The reduced pairing makes both of the states less bound,
resulting in a limited increase ($\sim$0.11 MeV)
in the calculated $E_x^{\textrm{th}}$ for the $1/2_1^-$ state.
It may be noted that while the WBT interaction provides a very good prediction
for the energy of $^{15}{\rm C}(1/2_1^-)$, this agreement
deteriorates for WBT(0.7$V_{01}$).

The locations of the cross-shell states in $^{15,17,19}{\rm C}$
are, as may be seen in Fig.~\ref{fig:migration_1hm_3hm_IIIb}, best described 
by the YSOX interaction predictions.   
Three favourable features of YSOX are worth noting in this context. 
First, the $\langle(d_{5/2})^2|V|(d_{5/2})^2\rangle_{J=0,T=1}$ pairing term 
has already been reduced
in the SDPF-M interaction~\cite{Yuan12}, on which the $sd$ part of YSOX is based. 
Second, the $T$=0 tensor force in the $p$-$sd$ cross-shell part of YSOX,
originating from the $\pi+\rho$ tensor force~\cite{Otsuka05},
is stronger than in the WBT interaction (see Fig.~2 (a) of Ref.~\cite{Yuan12}):
the monopole MEs 
$\langle p_{1/2}d_{5/2}|V|p_{1/2}d_{5/2} \rangle_{T=0}^{\rm T}$ 
($\langle p_{3/2}d_{5/2}|V|p_{3/2}d_{5/2} \rangle_{T=0}^{\rm T}$) 
are more attractive (repulsive) in YSOX. 
In closed $p$-shell nuclei, the overall
contribution from these two terms is limited, 
while in open $p$-shell nuclei, their interplay plays a key role. 
In the neutron-rich carbon isotopes of interest here, 
the effect is mainly repulsive -- 
the stronger repulsion between the $\pi p_{3/2}$ and $\nu d_{5/2}$ orbits 
raises the $\nu d_{5/2}$ orbit in energy, resulting in higher $E_x$ for the cross-shell states. 
Third, the monopole MEs
for the neutron-neutron ($T=1$) central force,
$\langle p_{1/2}d_{5/2}|V|p_{1/2}d_{5/2} \rangle_{T=1}^{\rm C}$ and
$\langle p_{3/2}d_{5/2}|V|p_{3/2}d_{5/2} \rangle_{T=1}^{\rm C}$,
are repulsive in WBT, while they are slightly attractive in YSOX 
(see Fig.~2 (b) of Ref.~\cite{Yuan12}).
This has the following consequences when one neutron 
in the $\nu p_{1/2}$ or $\nu p_{3/2}$ orbit 
is removed.  In the WBT description, 
the $d_{5/2}$ neutrons are subject to less repulsion 
from the neutrons in the $p$ orbits, 
resulting in smaller predicted $E_x^{\textrm{th}}$ for the hole states,
while for the YSOX interaction the opposite occurs. 
Note that in the neutron-rich carbon isotopes, 
the $\nu s_{1/2}$ and $\nu d_{5/2}$ orbits are essentially degenerated~\cite{Stanoiu08},
which leads to extensive configuration mixing owing to many-body correlations~\cite{Yuan12b}. 
As the $\nu d_{5/2}$ orbit plays a major role in the $sd$-valence space, 
the above argument will be valid for many of the excited states.

The MK interaction~\cite{Millener75} 
provides a moderately good description of the cross-shell states (Fig.~\ref{fig:migration_1hm_3hm_IIIb}).
This interaction shares a key character with YSOX in 
that it incorporates non-central components --
specifically the tensor force as fixed by the underlying $NN$ interaction. 
This feature is believed to be primarily responsible for the inversion of the neutron
$d_{5/2}$ and $s_{1/2}$ states in $^{15}{\rm C}$~\cite{Millener75}. 
It may be noted that the MK interaction was developed
to reproduce the spectroscopic features of nuclei
in the mass range $A$=11 to 16.  As shown here the extension to 
$^{17,19}{\rm C}$ suggests that it performs reasonably well at somewhat higher mass number.

In terms of the positive parity states it may be seen (Fig.~\ref{fig:spectroscopy_Sunji_17c_IV_pos})
that both the WBT and YSOX interactions
predict similar energies for the
$5/2_2^+$, $7/2_1^+$ and $9/2_1^+$ states, 
which compare well with experiment. 
On the other hand, the energies calculated using the WBT$^*$ interaction for these states 
are lower by $\sim$0.5 MeV. 
This is in contrast to the case of the $2^+_1$ states 
in $^{16,18,20}{\rm C}$~\cite{Stanoiu08,Petri12}, 
where WBT$^*$ better describes their locations. 
A similar observation may be made for $^{19}{\rm C}$ where the energy of
the $5/2^+_2$ state [$E_x$=1.46(10) MeV ~\cite{Ysatou08}] 
is well accounted for by WBT ($E_x^{\textrm{th}}$=1.40 MeV), 
but not by WBT$^*$ ($E_x^{\textrm{th}}$=1.08 MeV). 
Sieja \textit{et al.}~\cite{Sieja11} 
have shown that the $2^+_1$ energy in $^{16}{\rm C}$ 
could be satisfactorily reproduced within the shell model 
by introducing an asymmetric core ($^{10}{\rm He}$) 
and an effective interaction which takes into account 
proton core polarization contributions up to third order with the inclusion
of folded diagrams. 
This demonstrates that for asymmetric systems in which the valence spaces 
of protons and neutrons span two different major shells,
special attention needs to be paid to the construction of the shell-model MEs.
It may also be noted that the CCEI results~\cite{Jansen14}
provide a consistent description of the energies
for both the half-integer-spin $5/2_2^+$ and $9/2_1^+$ states in $^{17}{\rm C}$
(Fig.~\ref{fig:spectroscopy_Sunji_17c_IV_pos})
and the integer-spin $2^+_1$ states in $^{18,20}{\rm C}$ 
(see Fig.~3 of Ref.~\cite{Jansen14}). 
In $^{19}{\rm C}$ the energy of the $5/2^+_2$ state with respect to the $1/2_1^+$ level (experimentally the ground state)  
is also well reproduced by the CCEI calculations (Fig.~4 in Ref.~\cite{Hwang17}).

In summary, the spectroscopy of neutron unbound states in $^{17}{\rm C}$ has been investigated
using single-neutron removal from $^{18}{\rm C}$.
Resonances that were determined to lie at $E_x$=2.65(2) and 3.94(2)~MeV were demonstrated to correspond to $p$-wave hole states with $J^{\pi}$ of 
$1/2_1^-$ and $3/2_1^-$, respectively.  Additionally, another resonance at $E_x$=1.51(3)~MeV has been provisionally assigned $J^{\pi}$=$5/2_2^+$.  Given that this level can provide insight into the neutron-neutron interaction in proton-neutron
asymmetric systems~\cite{Stanoiu08} a confirmation of this assignment would be welcome.  Comparison with the predictions
provided by a range of shell-model interactions for levels in $^{17}{\rm C}$, as well as the neighbouring odd-$A$ isotopes  $^{15,19}{\rm C}$, demonstrated that the YSOX interaction provides the best agreement, including for the cross-shell
$1/2_1^-$ and $3/2_1^-$ states.

\begin{acknowledgments}
We are grateful to Dr.~C.~Yuan
for providing us with their shell-model Hamiltonian.
The present work was in part supported by 
the World Class University Project (R32-2008-000-10155-0), 
NRF grants (NRF-2011-0006492, NRF-2018R1A5A1025563)
and IBS (IBS-R031-D1, 2013M7A1A1075764:RISP) in Korea, 
and
JSPS KAKENHI Grant Numbers JP16H02179, JP18H05404, JP21H04465
and MEXT KAKENHI Grant Number 24105005 in Japan.
N.L.A., F.D., J.G., F.M.M. and N.A.O. acknowledge
partial support from the French-Japanese
LIA-International Associated Laboratory 
for Nuclear Structure Problems 
as well as the French ANR-14-CE33-0022-02 EXPAND. 
A.N. and J.G. would like to acknowledge the JSPS Invitation fellowship program
for long term research in Japan at the Tokyo Institute of Technology
and RIKEN, respectively.
S.L. gratefully acknowledges the support provided
by the RIKEN International Associate Program
and the hospitality of the Nishina Center Staff during his
sojourn. 
\end{acknowledgments}


\end{document}